\documentclass[aip,rsi,reprint]{revtex4-1}

\usepackage{graphicx}
\usepackage{amsmath}
\usepackage{array}
\usepackage{makecell}

\draft % marks overfull lines with a black rule on the right

\begin{document}

% Use the \preprint command to place your local institutional report number 
% on the title page in preprint mode.
% Multiple \preprint commands are allowed.
%\preprint{}

\title{A permanent-magnet Zeeman slower and magneto-optical trap for calcium atoms for ultracold Rydberg physics} %Title of paper

% repeat the \author .. \affiliation  etc. as needed
% \email, \thanks, \homepage, \altaffiliation all apply to the current author.
% Explanatory text should go in the []'s, 
% actual e-mail address or url should go in the {}'s for \email and \homepage.
% Please use the appropriate macro for the type of information

% \affiliation command applies to all authors since the last \affiliation command. 
% The \affiliation command should follow the other information.
\author{E. Marin-Bujedo}
\affiliation{Universit\'e catholique de Louvain, Institute of Condensed Matter and Nanosciences\\Chemin du Cyclotron 2, 1348 Louvain-la-Neuve, Belgium}

\author{J.A.L. Grondin}
\affiliation{Universit\'e catholique de Louvain, Institute of Condensed Matter and Nanosciences\\Chemin du Cyclotron 2, 1348 Louvain-la-Neuve, Belgium}
\affiliation{Katholieke Universiteit Leuven, Instituut voor Kern- en Stralingsfysica\\Celestijnenlaan 200, 3001 Heverlee, Belgium}

\author{T. Schiltz}
\affiliation{Universit\'e catholique de Louvain, Institute of Condensed Matter and Nanosciences\\Chemin du Cyclotron 2, 1348 Louvain-la-Neuve, Belgium}
\author{T. Corbo}
\affiliation{Universit\'e catholique de Louvain, Institute of Condensed Matter and Nanosciences\\Chemin du Cyclotron 2, 1348 Louvain-la-Neuve, Belgium}
\author{X. Urbain}
\affiliation{Universit\'e catholique de Louvain, Institute of Condensed Matter and Nanosciences\\Chemin du Cyclotron 2, 1348 Louvain-la-Neuve, Belgium}
\author{M. G\'en\'evriez}
\email[]{matthieu.genevriez@uclouvain.be}
%\homepage[]{Your web page}
%\thanks{}
%\altaffiliation{}
\affiliation{Universit\'e catholique de Louvain, Institute of Condensed Matter and Nanosciences\\Chemin du Cyclotron 2, 1348 Louvain-la-Neuve, Belgium}

% Collaboration name, if desired (requires use of superscriptaddress option in \documentclass). 
% \noaffiliation is required (may also be used with the \author command).
%\collaboration{}
%\noaffiliation

\date{\today}

\begin{abstract}
We report the construction and characterization of an experimental setup for producing a cold gas of $^{40}$Ca atoms and excite them to high Rydberg states with a resonant three-photon-excitation scheme.
The apparatus comprises four stages, each designed in-house. 
An oven heated to $\sim 500^\circ$C generates an atomic beam that is collimated by a capillary stack.
The beam is sent into a passive, permanent-magnet-based Zeeman slower that reduces the atomic velocity to $30$ {m/s}.
The slow atoms are captured in a magneto-optical trap (MOT) and cooled to $1.0(3)$ {mK} with a trapping time of $16(2)$ {ms}.
Ground-state atoms in the cold gas are excited to high Rydberg states via resonant excitation through the intermediate $4s4p\, ^1P_1$ and $4s4d\, ^1D_2$ states.
The MOT is operated at the center of an electrode stack, which serves to apply continuous and pulsed electric fields and field-ionize the Rydberg atoms for detection.
We benchmark our MOT against previous implementations and find its performance consistent with state-of-the-art results in terms of temperature and trapping lifetime.
Finally, we demonstrate Rydberg spectroscopy of calcium, confirming the system’s suitability for ultracold Rydberg physics experiments.
\end{abstract}

% insert suggested PACS numbers in braces on next line

\maketitle %\maketitle must follow title, authors, abstract and \pacs

% Body of paper goes here. Use proper sectioning commands. 
% References should be done using the \cite, \ref, and \label commands
\section{Introduction}
Magneto-optical traps (MOTs) have become an essential part of a large variety of experiments in atomic physics that require cold atoms for, e.g., high-precision spectroscopy \cite{kistersHighresolutionSpectroscopyLasercooled1994}, Bose-Einstein condensate \cite{andersonObservationBoseEinsteinCondensation1995}, atomic clocks \cite{wangMOTbasedContinuousCold2010, hinkleyAtomicClock10182013}, quantum simulation \cite{labuhnTunableTwodimensionalArrays2016}, and the study of highly excited atoms \cite{liMillimeterwaveSpectroscopyCold2003} and molecules \cite{collopy3DMagnetoOpticalTrap2018}.
While many atomic species have already been laser cooled in a MOT \cite{winelandRadiationPressureCoolingBound1978, phillipsLaserDecelerationAtomic1982, raabTrappingNeutralSodium1987, shumanLaserCoolingDiatomic2010, baum1DMagnetoOpticalTrap2020, vilasMagnetoopticalTrappingSubDoppler2022}, recent efforts have focused on alkaline-earth atoms.
Because they possess two valence electrons, such atoms exhibit an energy-level structure that is both relatively simple compared to most atomic species, yet more complex than that of the widely used alkali atoms.
This gives access to a broader range of physical phenomena and ways to control and manipulate the atoms.
For example, the coupling between the spins of the two valence electrons, together with the Pauli exclusion principle, leads to the formation of singlet and triplet states.
These states give rise to metastable excited levels that are crucial for metrological applications such as optical atomic clocks \cite{udemAbsoluteFrequencyMeasurements2001, yinDevelopmentStrontiumOptical2018}.
Additionally, the presence of narrow intercombination lines allows the atoms to reach much lower temperatures \cite{degenhardtCalciumOpticalFrequency2005}.
Finally, the presence of two electrons allows one to manipulate \cite{wilsonTrappingAlkalineEarth2022, burgersControllingRydbergExcitations2022, phamCoherentLightShift2022a, wirthQuadrupoleCouplingCircular2024}, directly laser cool \cite{bouillonDirectLaserCooling2024, lachaudSlowingCoherentSuperposition2024}, or interrogate \cite{muniOpticalCoherentManipulation2022a} Rydberg atoms in a nondestructive manner. 

\begin{figure}[h!]
    \centering
    \includegraphics[width=0.4\textwidth]{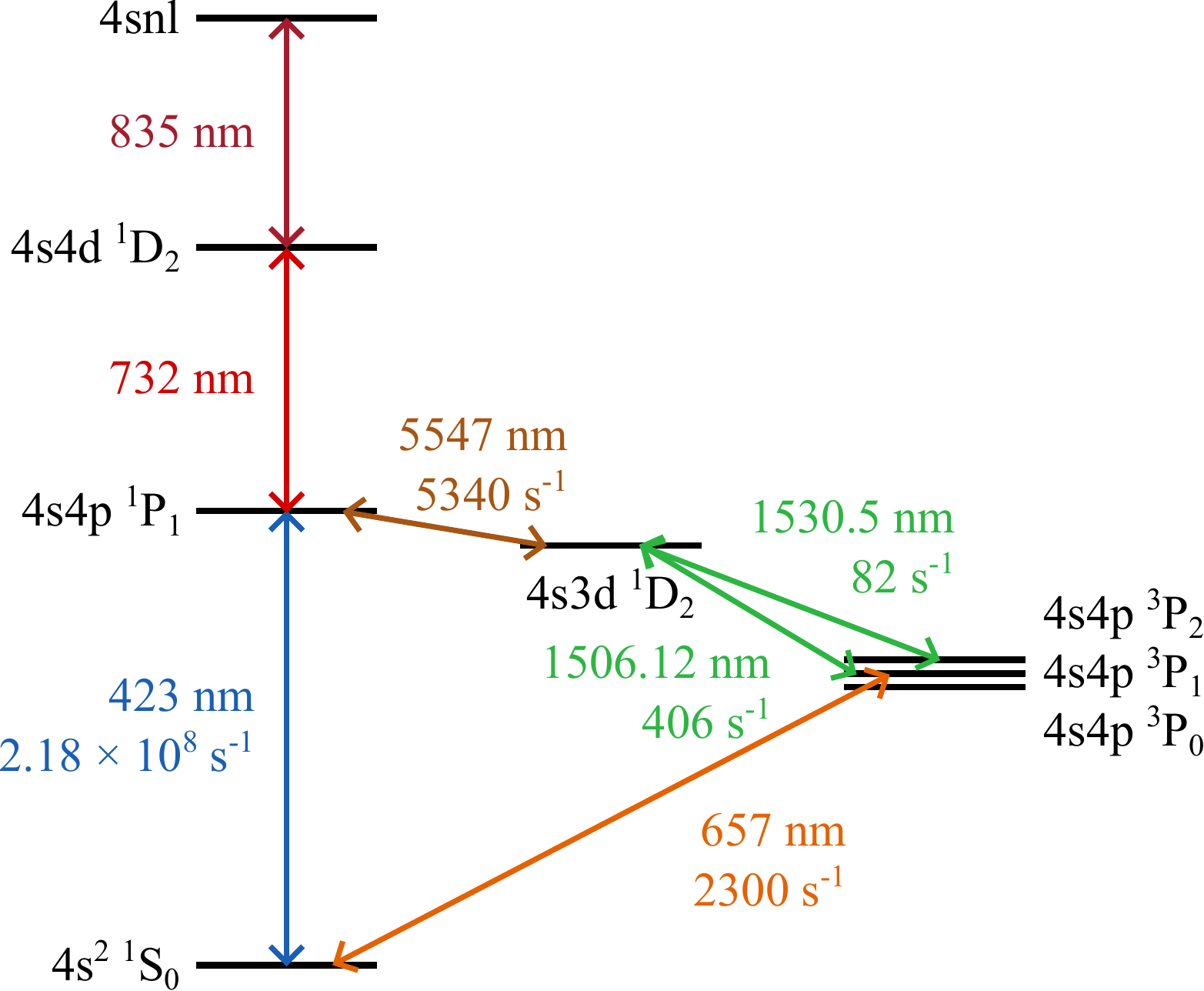}
    \caption{Relevant energy levels and transitions of $^{40}$Ca. Transition wavelengths were obtained from Ref. \cite{AtomicSpectraDatabase2009}. Einstein $A$ coefficients were obtained from Ref. \cite{adamczykTwophotonCoolingCalcium2025, wilpersAbsoluteFrequencyMeasurement2007}.}
    \label{fig:energy level scheme}
\end{figure}

The Calcium atom is an advantageous candidate for exploring the applications in Rydberg physics given above, in particular because the autoionization rates of its doubly excited Rydberg states are low \cite{marin-bujedoAutoionizationHighCoreexcited2023}.
Ultracold Ca Rydberg gases have been much less explored than those of heavier species (Sr \cite{lachaudSlowingCoherentSuperposition2024, muniOpticalCoherentManipulation2022a}, Yb \cite{wilsonTrappingAlkalineEarth2022, hinkleyAtomicClock10182013, phamCoherentLightShift2022a}), despite their energy level scheme being well suited for laser cooling and Rydberg excitation (see Fig. \ref{fig:energy level scheme}).
The frequency of the primary cooling transition Ca($4s^2\, ^1S_0$ – $4s4p\, ^1P_1$) lies in the visible range ($423$ {nm}), which is conveniently accessed with standard external cavity diode lasers, as used in our setup.
Moreover, the broad linewidth of this transition ($34$ {MHz}) makes it ideal for scattering a large number of photons and cooling down the atom efficiently.
In addition $^{40}$Ca, the most abundant isotope of Ca with $96.64(16)\%$  abundance \cite{salumbidesHighprecisionFrequencyMeasurement2011}, has zero total nuclear spin, and hence no hyperfine structure, which further simplifies the cooling scheme.

Calcium laser cooling was first reported in 1991, when one-dimensional optical molasses were used to deflect the slow atoms coming out of a Zeeman slower \cite{witteLaserCoolingDeflection1992}.
A few years later, the first calcium MOT was demonstrated by the same group \cite{kistersHighresolutionSpectroscopyLasercooled1994}.
Since then, a number of Ca MOTs have been constructed \cite{kurosuLaserCoolingTrapping1990, grunertSubDopplerMagnetoopticalTrap2002, cavassofilhoCalciumMagnetoopticalTrap2003, dammalapatiCompactMagnetoopticalTrap2009, burrowsRepumpMethodsMagneto2012, millsEfficientRepumpingCa2017}, achieving typical number of atoms of $10^6 - 10^8$, and temperatures between $1.3$ {mK} \cite{kistersHighresolutionSpectroscopyLasercooled1994} and $9$ {mK} \cite{cavassofilhoCalciumMagnetoopticalTrap2003}.
The most pronounced differences among reported Ca MOTs lie in their trapping times, which span from $16$ {ms}\cite{burrowsRepumpMethodsMagneto2012} to $2.5$ {s}\cite{millsEfficientRepumpingCa2017}.
This large variability is primarily due to whether repumping systems are employed or not.
The upper cooling state $4s4p\, ^1P_1$ can leak through $4s3d\, ^1D_2$ to the triplet states $^3P_2$ and $^3P_1$ (see Fig. \ref{fig:energy level scheme}), with a branching ratio of $1:10^5\,$ \cite{beveriniMeasurementCalcium1P11D21989}.
Although the ratio is sufficiently favourable for the atoms to be laser cooled, this decay pathway represents the main limiting factor to the MOT lifetime.
Without any repumping, the trapping time is limited to $\sim 16$ {ms}\cite{burrowsRepumpMethodsMagneto2012}.
Adding a $672$ {nm} repump laser to drive the Ca($4s3d\, ^1D_2$ – $4s5p\, ^1P_1$) transition increases the lifetime by a factor of three \cite{burrowsRepumpMethodsMagneto2012}.
Further extension of the trapping time can be achieved either by choosing another repump wavelength\cite{millsEfficientRepumpingCa2017} or by adding a second repump laser\cite{burrowsRepumpMethodsMagneto2012}.
Other parameters, such as beam size, available laser power, laser detuning, and optical alignment, can also influence the trapping time, number of trapped atoms, and/or final temperature \cite{dammalapatiCompactMagnetoopticalTrap2009, adamczykTwophotonCoolingCalcium2025, cavassofilhoCalciumMagnetoopticalTrap2003}.

In the last years, more advanced cooling schemes have been developed to bring the Ca MOT temperature well below its Doppler limit of $0.8$ {mK}.
One example is the two-color MOT \cite{adamczykTwophotonCoolingCalcium2025, degenhardtCalciumOpticalFrequency2005,grunertSubDopplerMagnetoopticalTrap2002}, where the trap is first loaded using the broad $423$-nm transition to capture a large number of atoms even with relatively large velocities, and a second, narrower-linewidth transition is used in a second stage to cool the atoms to lower temperatures.

The present paper reports the construction and characterization of a magneto-optical trap setup using the $423$-nm laser cooling transition without any repumper.
The magneto-optical trap is loaded with slow atoms coming out of a Zeeman slower, itself loaded with atoms from a Calcium oven.
Each system is described in detail in Sec. \ref{sec:experimental setup} (\ref{subsec:overview}-\ref{subsec:magneto-optical trap}).
The cold atoms in the MOT are then excited to high Rydberg states using the three-photon excitation scheme shown in Fig. \ref{fig:energy level scheme}.
The scheme was used with thermal beams \cite{miyabeDeterminationIonizationPotential2006, nortershauserLineShapesTripleresonance2000} but not in Ca MOTs, which focused on two-photon schemes \cite{zelenerMeasurementsRydbergTransition2019}.
The Rydberg atoms are then detected using a segmented electrode stack described in Sec. \ref{subsec:electrode stack}.
The setup is characterized in Sec. \ref{sec:results}, with particular emphasis on the performances of the trap.
We are finally discussing Rydberg excitation of the cold atomic cloud in the MOT, demonstrating the possibility of using the present setup to investigate ultracold gases of Ca Rydberg atoms.

\section{Experimental setup}
\label{sec:experimental setup}
\subsection{Overview of the experiment}
\label{subsec:overview}
\begin{figure*}[t!]
    \centering
    \includegraphics[width=0.8\textwidth]{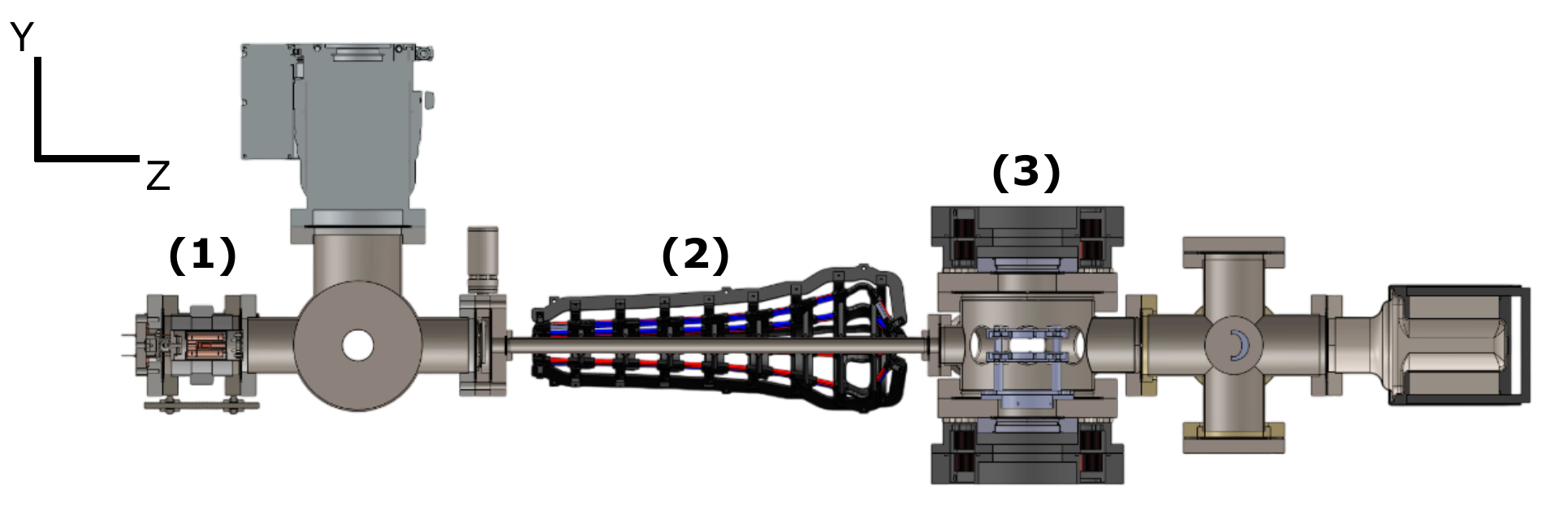}
    \caption{Sectional view of the experimental setup, where \textbf{(1)} is the Ca oven, \textbf{(2)} is the permanent-magnet-based Zeeman slower, and \textbf{(3)} is the magneto-optical trap.}
    \label{fig:setup side-view half-cut}
\end{figure*}

Figure \ref{fig:setup side-view half-cut} presents a sectional view of the experiment setup 3D model. A single $423$ {nm} external cavity diode laser (ECDL), frequency-locked to the Ca($4s^2\, ^1S_0\; -\; 4s4p\, ^1P_1$) atomic transition, is employed for slowing, cooling, and trapping calcium atoms under ultra-high vacuum conditions ($< 10^{-8}$ {mbar}).
The atoms are thermally emitted from a custom-built oven (Fig. \ref{fig:setup side-view half-cut}\textbf{(1)}) operating at $500${$^o$C}, initially traveling at approximately $650$ {m/s}.
They are decelerated by a counter-propagating laser beam within a permanent-magnet Zeeman slower (Fig. \ref{fig:setup side-view half-cut}\textbf{(2)}) to velocities below the capture threshold of the magneto-optical trap.
The slowed atoms then enter the central region of the MOT chamber (Fig. \ref{fig:setup side-view half-cut}\textbf{(3)}), where three orthogonal pairs of circularly polarized laser beams intersect.
A pair of electromagnetic coils in anti-Helmholtz configuration on the top and bottom of the MOT chamber generates the required magnetic field gradient at the center of the trap.
The cold atom cloud is observed via fluorescence using a CMOS camera and a photomultiplier tube.

\subsection{Calcium oven}
\label{subsec:oven}
\begin{figure}[h!]
    \centering
    \includegraphics[width=0.5\textwidth]{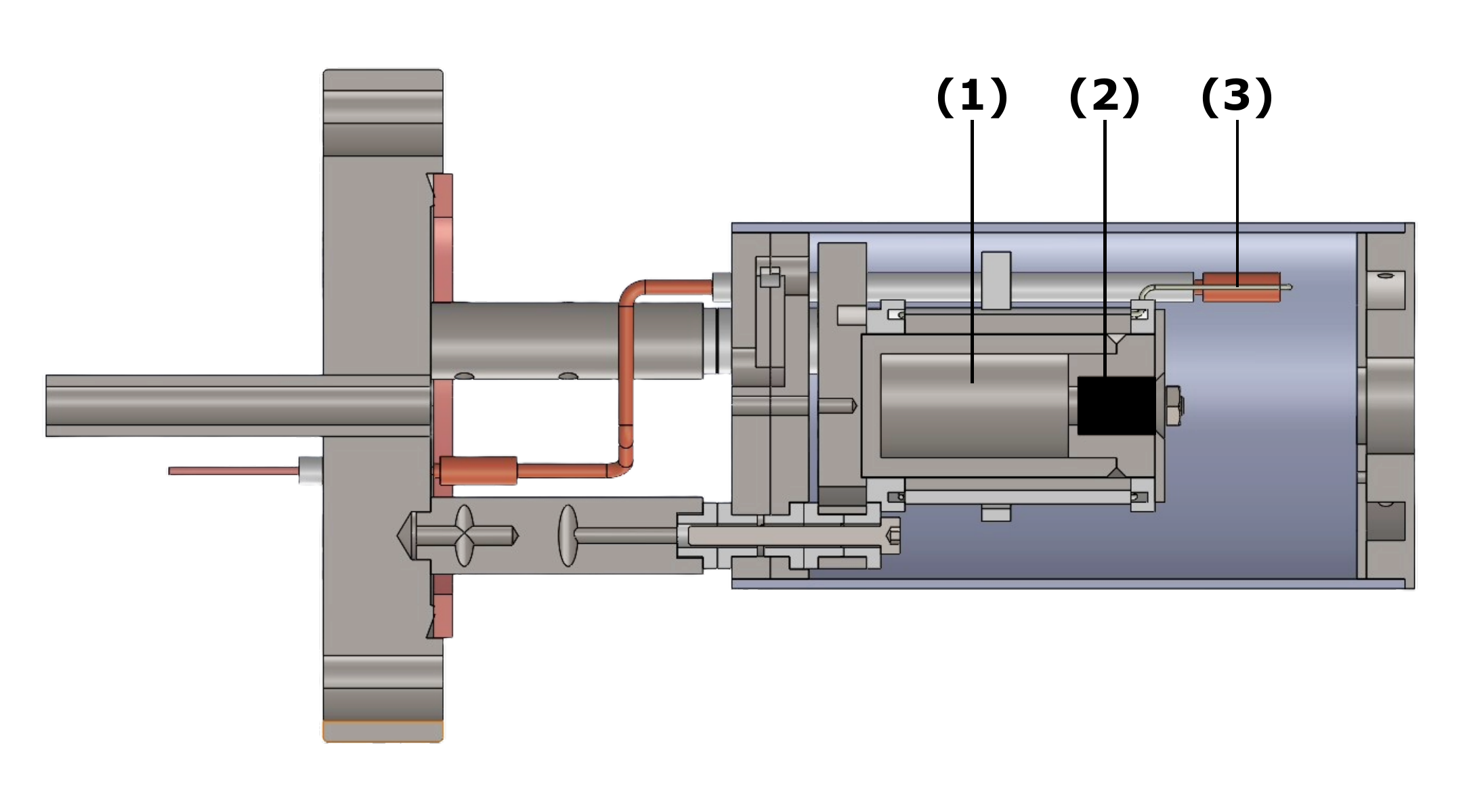}
    \caption{Sectional view of the Ca oven. \textbf{(1)} is the cartridge where the Ca granules are loaded, \textbf{(2)} are the capillary tubes, and \textbf{(3)} is the tantalum wire.}
    \label{fig:oven}
\end{figure}

The oven was built based on the design of Schioppo et al. \cite{schioppoCompactEfficientStrontium2012}.
It consists of a cylindrical chamber with a tantalum wire acting as the heating element, arranged around the lateral surface (Fig. \ref{fig:oven}).
Calcium, in the form of small granules, is loaded into a cylindrical cartridge that is then inserted into the oven.
The oven output is passing through capillary tubes with a radius of $R = 0.15$ mm and a length of $L = 12$ mm.
The aspect ratio of the capillaries, $L/R$, determines the divergence of the atomic beam and, together with the temperature, the atomic flux.
The measured atomic-beam divergence is $0.11$ {rad}, corresponding to a transverse velocity of $60$ {m/s}, or equivalently $141$ {MHz} Doppler broadening (FWHM).
It is larger than the divergence that would be expected for a single capillary with $L/R=80$ ($\sim  0.04$ {rad}), most likely because of the imperfect alignment of the capillaries within the stack.
The oven is mounted on a CF$40$ flange attached to the experiment’s vacuum chamber and is surrounded by a thermal shield that reduces energy losses, enabling temperatures of approximately $500^\circ${C} necessary to evaporate sufficient calcium atoms \cite{norrisLaserCoolingTrapping} with an input electrical power of 20 {W}.
The atomic flux is estimated, from the resonant absorption of 423-nm light, to be $10^{14}$ {cm$^{-2}$s$^{-1}$}.

\subsection{Light production and control}
\label{subsec:light control}
\begin{figure}[h!]
    \centering
    \includegraphics[width=0.5\textwidth]{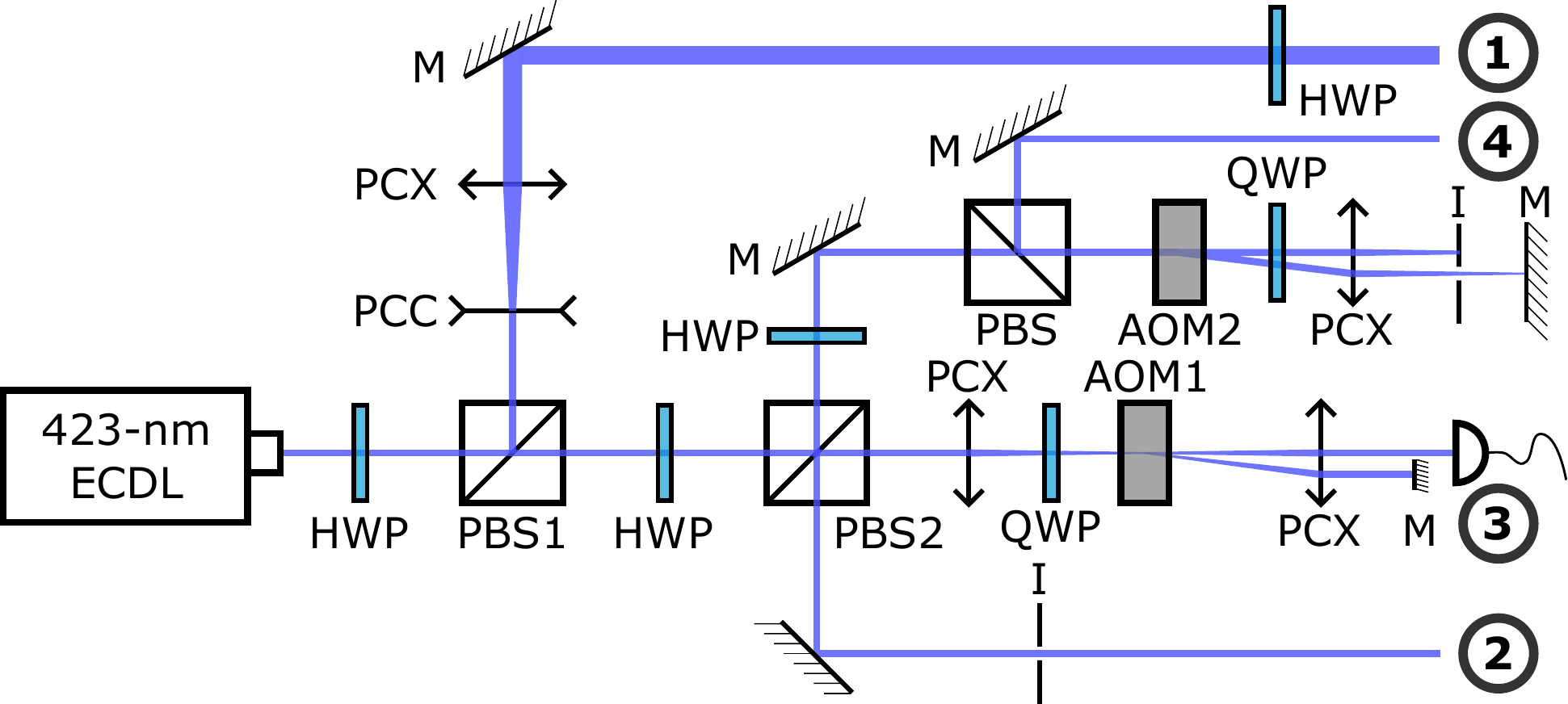}
    \caption{Optical path schematic. \textbf{HWP}: half-wave plate. \textbf{QWP}: quarter-wave plate. \textbf{PBS}: polarizing beam splitter. \textbf{PCX}: plano-convex lens. \textbf{PCC}: plano-concave lens. \textbf{AOM}: acousto-optic modulator. \textbf{M}: mirror. \textbf{I}: iris. The optical paths $1-4$ correspond to the slowing beam, frequency stabilization, beam diagnosis, and MOT beam, respectively.}
    \label{fig:optical path}
\end{figure}

Figure \ref{fig:optical path} shows a schematic view of the optical path.
The Ca cooling and trapping is driven by a single external cavity diode laser (ECDL) at $423$ {nm} \footnote{MOGLABS LDL Littrow enhanced ECDL}, delivering a total optical power of $84$ {mW}. This output is split into four optical paths, with the relative power in each path controlled by half-wave plates (HWPs) and polarizing beam splitters (PBSs).

The beam reflected from PBS1 is expanded with a Galilean telescope to a waist of {1.5} mm. This path (\textbf{1}) delivers $32$ {mW} of optical power and serves as the Zeeman slowing beam.

The beam transmitted through PBS2 is sent through a $300$-{MHz} acousto-optic modulator (AOM1). The first diffraction order is retroreflected in a cat’s-eye configuration \cite{donleyDoublepassAcoustoopticModulator2005}, resulting in a double pass through the modulator and a net frequency upshift of $600$ {MHz}.
This beam, sent into (\textbf{2}), carries $1$ {mW} and is used for frequency stabilization. The zeroth diffraction order from AOM1 constitutes beam (\textbf{3}), with $4$ {mW} power, and is used for diagnostic purposes.

The beam reflected on PBS2 is directed to the acousto-optic modulator AOM2, also operated in a double-pass cat’s-eye configuration with a tunable drive frequency. The resulting output (\textbf{4}) is upshifted by $600 + \Delta\nu_{\textrm{MOT}}$ {MHz}, where $\Delta\nu_{\textrm{MOT}} = -36$ {MHz} is the MOT detuning. This path (\textbf{4}) is provided with $8$ {mW} of optical power. The beam is subsequently split into three paths to form the MOT beams, each expanded to $2$ {mm} waist with a telescope.

\subsection{Zeeman slower}
\label{subsec:Zeeman slower}
Zeeman slowers utilize the Zeeman effect to compensate for the decreasing Doppler shift as the atoms decelerate, maintaining resonance between the atoms and the slowing laser along the entire length of the slower. The ideal field profile along the atomic beam axis is described by \cite{cheineyZeemanSlowerDesign2011},

\begin{equation}
B(z) = \frac{\eta \hbar k}{g_L \mu_B} \left[ \sqrt{v_0^2 - \frac{\left( v_0^2 - v_f^2 \right)}{L}z} - \frac{\delta}{k} \right]
\label{eq:zeeman profile}
\end{equation}

where $\eta$ is a design parameter, $k$ is the light wavenumber, $\mu_B$ is the Bohr magneton, $g_L = 1$ is the Land\'e factor of the upper state (the ground state is not Zeeman shifted), $L$ is the Zeeman slower's length, $\delta$ is the laser detuning and $v_0$ and $v_f$, the maximal velocity that can be slowed and the exit velocity, respectively.

Conventional Zeeman slowers often employ solenoidal coils to generate the required magnetic field profile \cite{phillipsLaserDecelerationAtomic1982}. Such systems require power supplies and water cooling due to the ohmic heating.
To overcome that issue, we implemented a passive Zeeman slower based on permanent magnets \cite{cheineyZeemanSlowerDesign2011}. 

In solenoid-based Zeeman slowers, the magnetic field is oriented along the direction of the atomic beam.
Circularly polarized light ($\sigma^+$ or $\sigma^-$) is required to drive the transition between the ground-state magnetic sublevel $m = 0$ and the excited-state sublevels $m = \pm 1$.
In contrast, in typical permanent-magnet designs, the magnetic field is oriented perpendicularly to the atomic beam.
Driving the same transitions in this scheme is only possible with linearly polarized light perpendicular to the magnetic field, and since the magnetic sublevels $m = +1$ and $m = -1$ are equally coupled, only $50\%$ of the total laser power is used to drive the right slowing transition.

\begin{figure}[h!]
    \centering
    \includegraphics[width=0.5\textwidth]{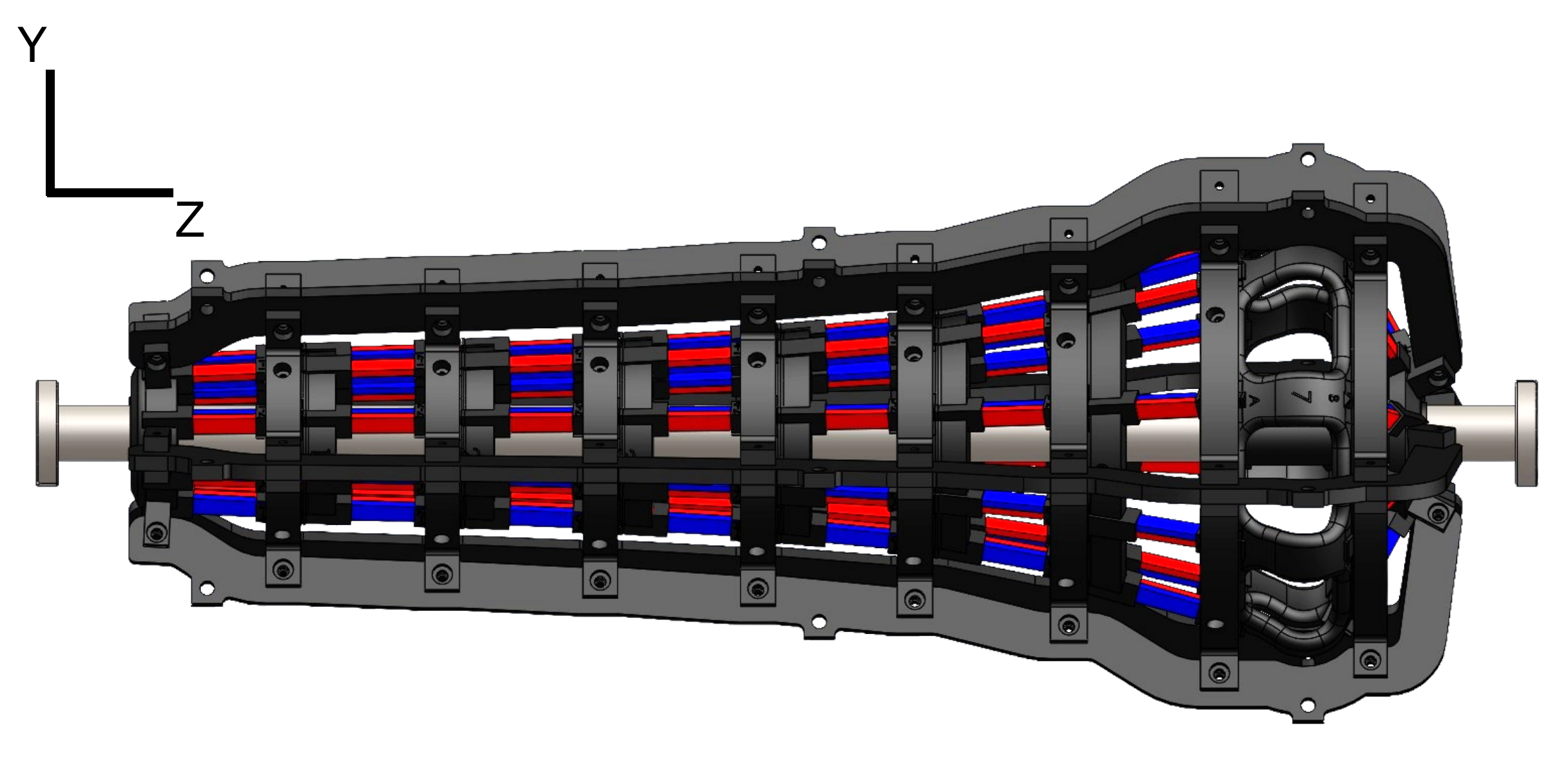}
    \caption{Zeeman slower 3D model. The red and blue segments represent the north and south poles of the magnets, respectively. The black elements are the $3$D-printed holding structure.}
    \label{fig:zeeman 3d}
\end{figure}

\begin{figure}[h!]
    \centering
    \includegraphics[width=0.5\textwidth]{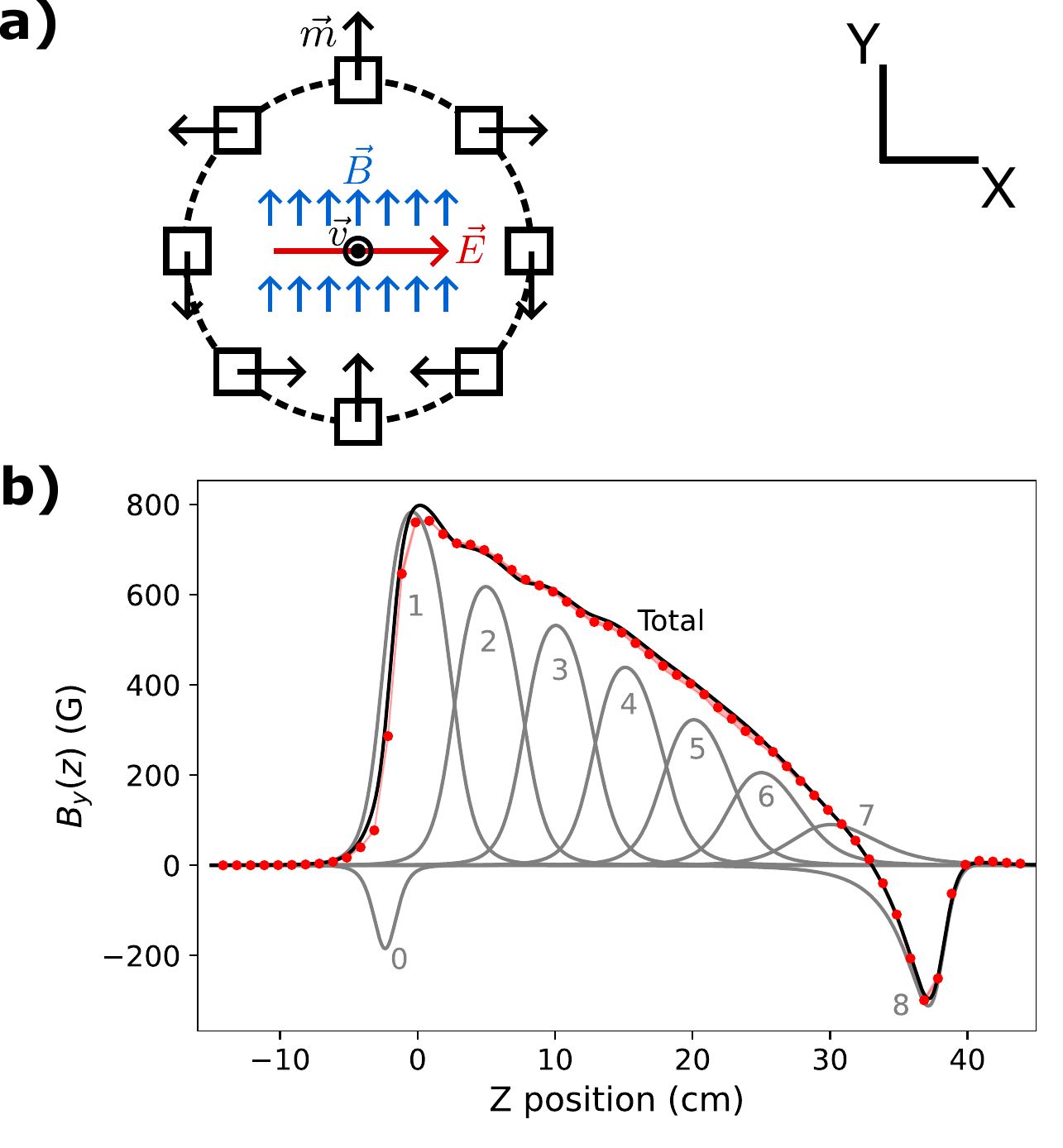}
    \caption{\textbf{a)} Radial arrangement of the magnets in Halbach configuration. The black, blue and red arrows represent the magnetic moment of the magnets, the homogeneous transverse magnetic field and the laser polarization, respectively. \textbf{b)} Longitudinal magnetic field profile. The field was calculated from the field generated by each magnet using the formulas given in \cite{cheineyZeemanSlowerDesign2011}. The contribution due to each segment of magnets is represented by gray solid lines numbered from $0$ to $8$, and the total field is represented by the black solid line. The red, filled circles and shaded area correspond to the experimental measurement of the magnetic field and its uncertainty, estimated as the standard deviation of three different measurements.}
    \label{fig:zeeman fields}
\end{figure}

As shown in figure \ref{fig:zeeman 3d}, the design comprises eight segments, each containing eight NdFeB cuboid magnets ($50 \times 6 \times 5$ {mm})\footnote{Magnets manufactured by HKCM engineering - REF: 9964-1799}, and another set of eight small magnets ($10 \times 3 \times 2$ {mm})\footnote{Magnets manufactured by HKCM engineering - REF: 9964-61967}.
The magnets are radially arranged in Halbach configuration \cite{halbachDesignPermanentMultipole1980}, which produces a homogeneous field in the transverse section (fig. \ref{fig:zeeman fields}\textbf{a}) while generating the required longitudinal profile (fig. \ref{fig:zeeman fields}\textbf{b}).
A custom-made $3$D-printed structure holds the magnets in position and ensures precise alignment of the slower with the atomic beam.
The $3$D-printed structure, fabricated from polylactic acid (PLA), can be printed within a day on regular printers, and it is sufficiently rigid to hold the magnets in place for over 2 years.
Moreover, the structure was designed such that magnet positioning is accurate to within $1$ {mm} in the radial and longitudinal positions, and $0.5 ^\circ$ in angle, ensuring that the actual configuration matches well the simulated one.
With this design, the Zeeman slower can be assembled within a few hours and at a very small cost compared to solenoid-based designs.
In this scheme, the atomic beam, magnetic field, and laser polarization are mutually orthogonal, as can be seen in fig. \ref{fig:zeeman fields}\textbf{a}.

The magnetic field, ranging from $800$ {G} at the entrance of the slower to $-250$ {G} at the exit, along with the laser detuning ($\delta = -600$ {MHz}), allows the slowing of atoms moving at up to $720$ {m/s}, corresponding to $64\%$ of the thermal distribution at $500$ {$^o$C}.
The magnetic field was measured experimentally twice, one year apart, yielding the same results without any adjustment of the $3$D-printed structure (Fig. \ref{fig:zeeman fields}\textbf{b}, red solid circles).

\begin{figure}[h!]
    \centering
    \includegraphics[width=0.5\textwidth]{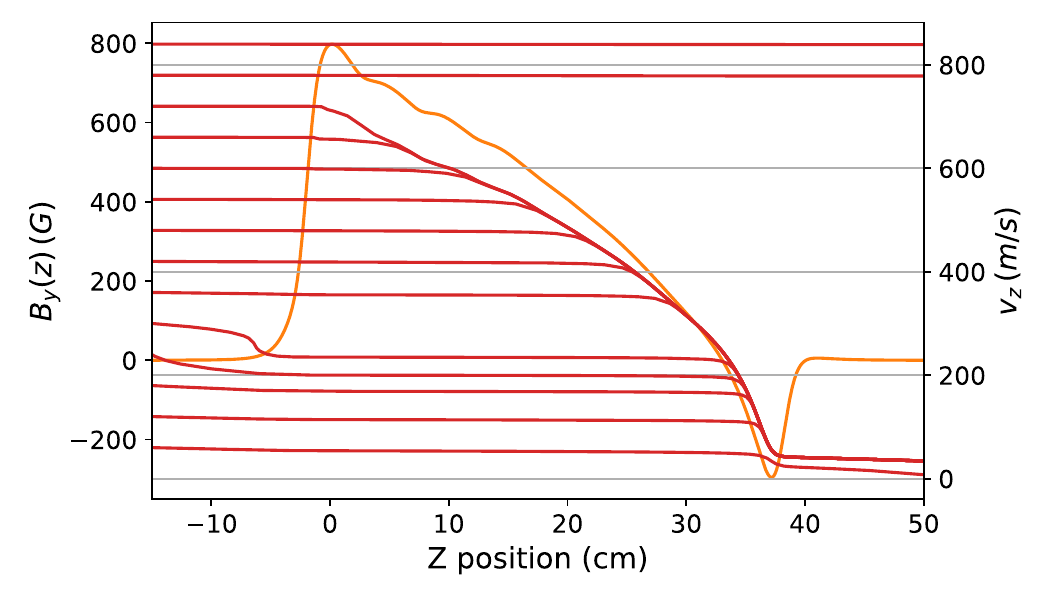}
    \caption{Red lines: calculated trajectories of the atoms in the phase space of longitudinal position ($z$) and longitudinal velocity ($v_z$) as they travel across the Zeeman slower. Orange line: $y$ component of the magnetic field on the axis of the Zeeman slower.}
    \label{fig:zeeman trajectories}
\end{figure}

The length of the Zeeman slower is set to ensure that the atoms are decelerated to a velocity of $\sim 35$ {m/s}, without being completely stopped at its end.
The acceleration considered to calculate the length of the slower is scaled by $\eta$ so that $a = \eta \cdot a_{\textrm{max}}$, where $\eta = 0.5$ and $a_{\textrm{max}} = \Gamma \hbar k / 2 m$ \cite{cheineyZeemanSlowerDesign2011} is the maximal acceleration exerted by the light onto the atoms.
Simulations of the trajectories of the atoms for different laser parameters indicated that a beam radius of $1.5$ {mm} (at $32$ {mW} total laser power) maximizes the number of atoms effectively slowed (Fig. \ref{fig:zeeman trajectories}).

On the one hand, the present, permanent-magnet-based, passive design offers less flexibility than the solenoid-based counterpart, since it does not permit field adjustment via current tuning.
Furthermore, only $50\%$ of the laser power is effectively used for slowing the atoms.
On the other hand, it eliminates any power consumption and thus, it avoids the necessity of any cooling system to prevent overheating.
In addition, the simulation shows a rapid decay of the magnetic-field magnitude outside the Zeeman slower, reaching a value well below the Earth's magnetic field ($\sim 0.5$ {G}) at the center of the MOT region located further downstream.

\subsection{Magneto-optical trap}
\label{subsec:magneto-optical trap}
Magneto-optical traps combine Doppler cooling and the Zeeman effect to cool and trap neutral atoms \cite{raabTrappingNeutralSodium1987}.
They consist of three orthogonal pairs of counter-propagating laser beams with opposite circular polarizations intersecting at the center of a quadrupole magnetic field generated by two coils in anti-Helmholtz configuration.

In our setup, the center of the MOT is located approximately $12$ {cm} downstream from the end of the Zeeman slower.
We operate the MOT using a $423$-nm laser red-detuned by $-36$ {MHz} from resonance.
The beam is split into three using polarizing beam splitters and half-wave plates, which allow independent control of the power delivered to each arm.
A quarter-wave plate is placed on each arm before entering the MOT chamber to convert the polarization of the beams from linear to circular.
After the chamber, each beam is retroreflected using a mirror and an additional quarter-wave plate, which reverses the direction of the circular polarization and creates standing waves along all three spatial axes.

\begin{figure}[h!]
    \centering
    \includegraphics[width=0.3\textwidth]{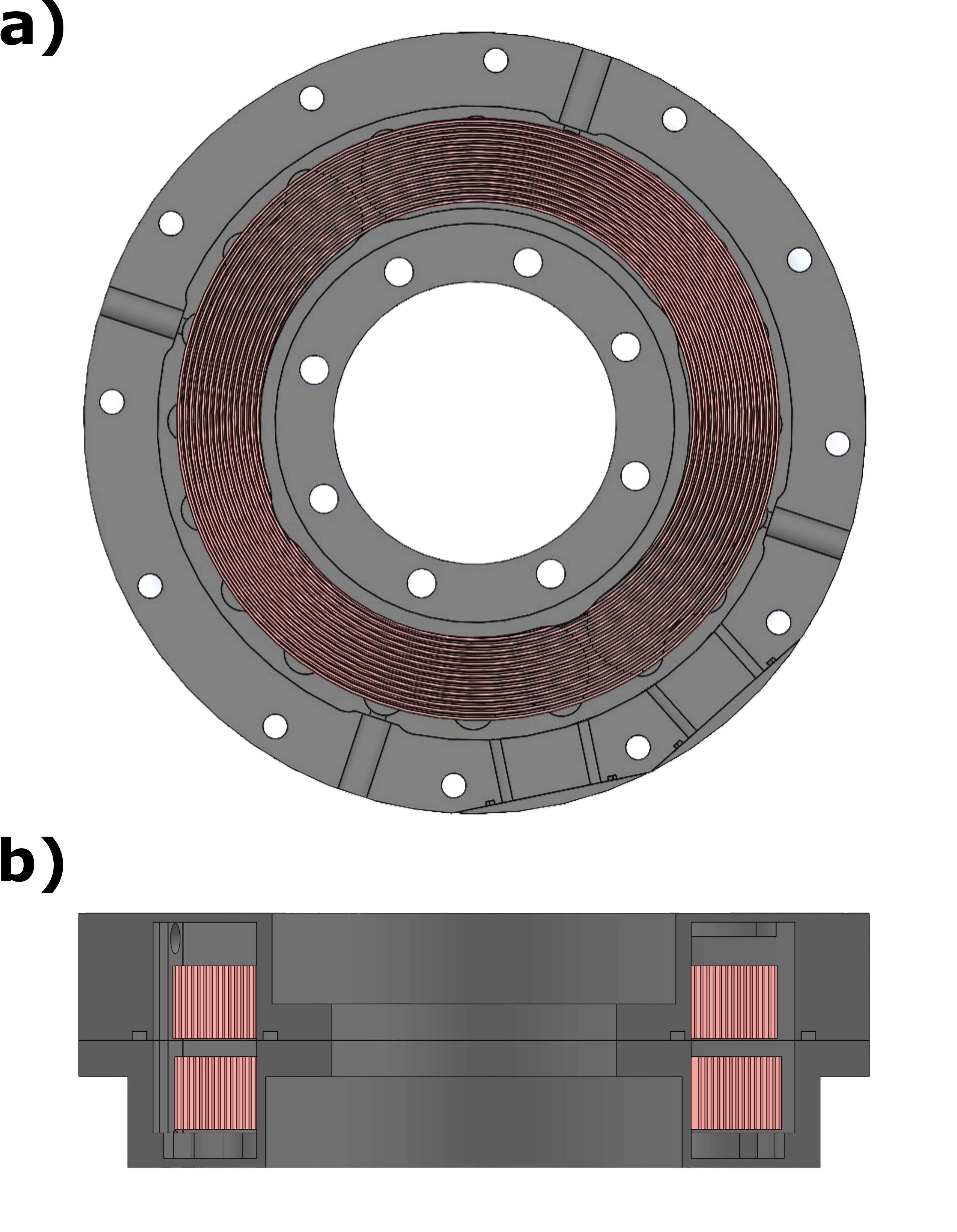}
    \caption{Top-view \textbf{(a)} and cut-half side-view \textbf{(b)} of the MOT coils.}
    \label{fig:coils}
\end{figure}

\begin{figure}[h!]
    \centering
    \includegraphics[width=0.4\textwidth]{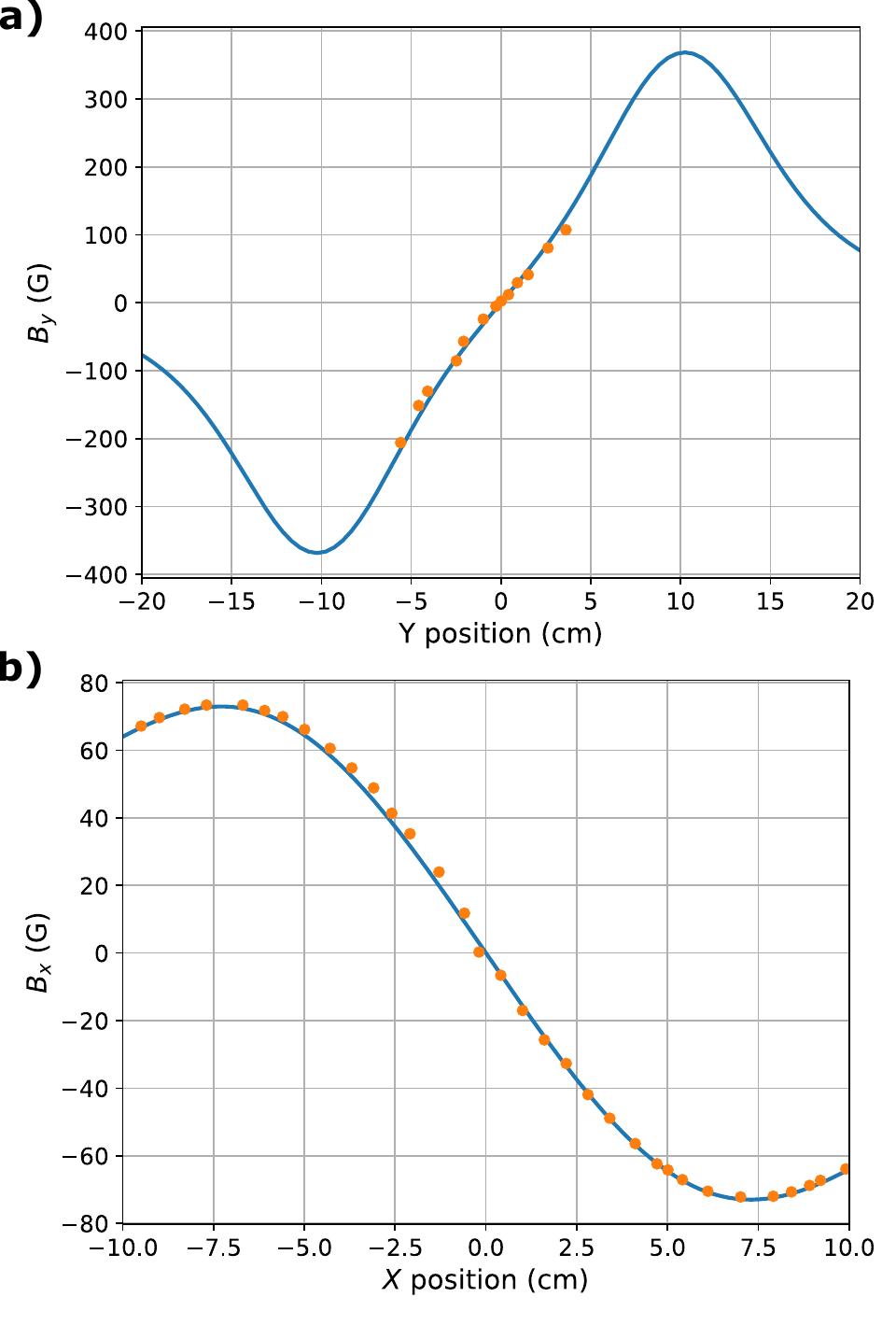}
    \caption{Magnetic field along the axial \textbf{a)} and radial \textbf{b)} directions of the coils. The blue line represents the calculated field, and the orange solid circles, the experimental measurements.}
    \label{fig:MOT - fields}
\end{figure}

The quadrupole magnetic field required for confinement is generated by a pair of coils in a quasi-anti-Helmholtz configuration, positioned above and below the MOT chamber.
The coils were custom-built in-house from a copper ribbon ($20$ {mm} wide, $0.7$ {mm} thick).
Each coil consists of two vertically stacked spirals with $15$ turns each, having internal and external diameters of $118$ {mm} and $168$ {mm}, respectively (Fig. \ref{fig:coils}).
The turns are spaced by $0.5$ {mm} with pieces of electric tape, and the two spirals are separated by $10$ {mm}.
With a current of $150$ {A} at $6$ {V}, the coils produce a magnetic field gradient of $30$ {G/cm} along the axial direction and $12$ {G/cm} in the radial directions.
The magnetic field experimentally measured along both the radial and the axial directions of the coils falls in very good agreement with the calculations, as shown in Fig. \ref{fig:MOT - fields}

To dissipate the heat produced by the high current circulating in the copper ribbon ($\sim 1$ {kW}), the coils were enclosed in custom-designed, watertight housings fabricated in-house from polyvinyl chloride (PVC) (see Fig. \ref{fig:coils}).
Deionized water circulates in the $0.5$-mm spacing between the coil windings, which is left free apart from a few spacing elements, ensuring an efficient heat dissipation due to the large contact area between water and copper, thanks to the ribbon shape of the conductor.
These housings are connected in parallel via $6$ mm diameter water pipes to a water chiller, which continuously circulates the deionized water through them to maintain the temperature at $25^\circ${C}.
The present design does not require hollow conductors in which water is passed through a hole in the center of the wire.
The conductance of the cooling system is large, such that the pressure drop across the coils is small, and water chillers with relatively small pumps can be used.
The simplicity of the setup means it can be wound manually, which makes it attractive for coils that require a small number of windings, and its high heat dissipation capabilities means the coils can be operated at large currents.
We have operated on a regular basis the coils up to $220$ {A}, the limit of our power supply, without overheating.

\begin{figure}[h!]
    \centering
    \includegraphics[width=0.3\textwidth]{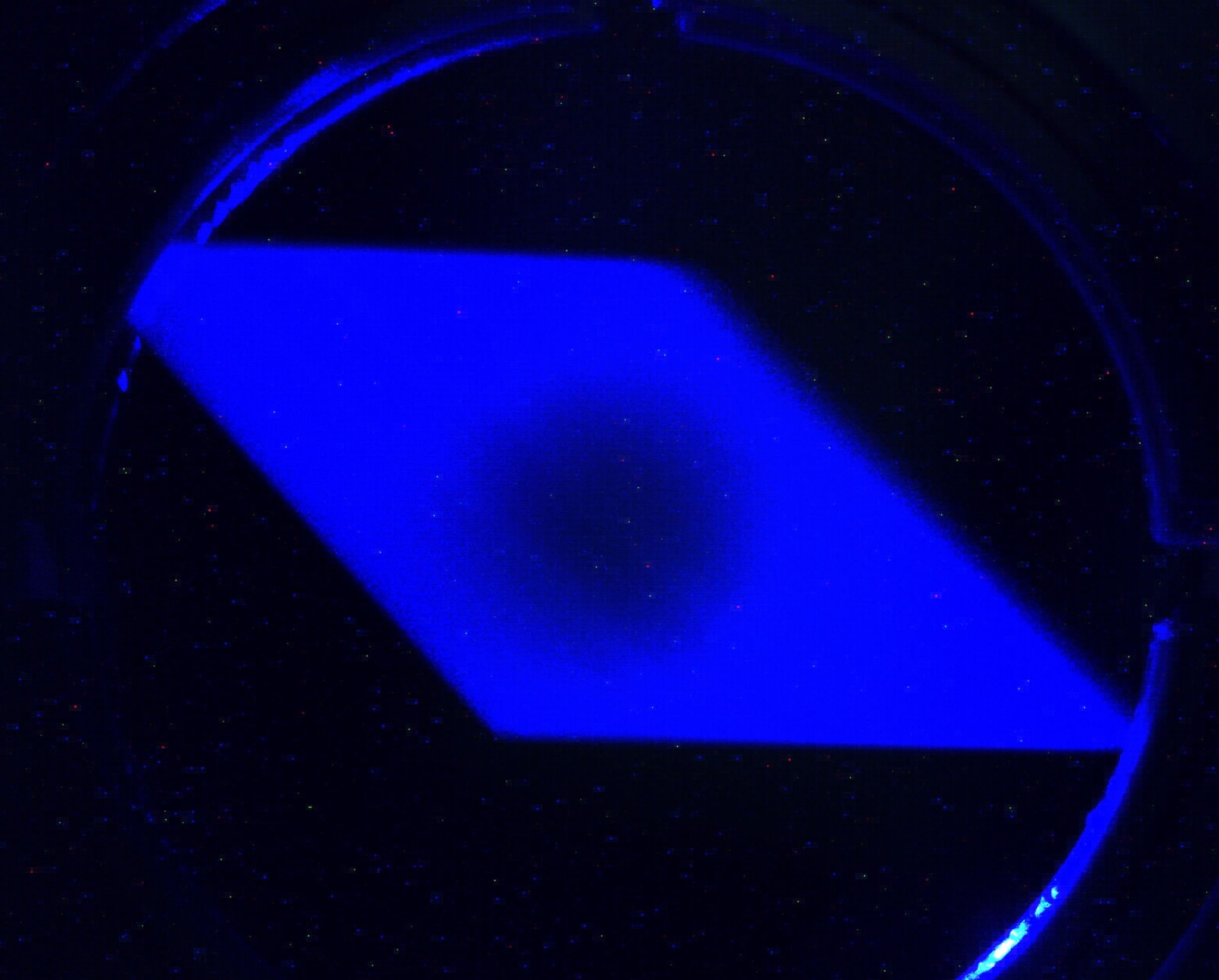}
    \caption{Zero Magnetic field position (dark area) inside the MOT observed by using the directional Hanle effect. The image was generated by combining the frames of a video recorded while sweeping a $423$-{nm} laser across the MOT region, and subsequently editing it to visualize only the brightest pixels, which gives the laser beam the rhombus-like shape observed in the picture. The edges of the $63$-CF laser window placed on top of the MOT chamber are also visible.}
    \label{fig:MOT - zero field position}
\end{figure}

To ensure that the MOT laser beams cross at the exact position where the magnetic field is zero, we used the directional Hanle effect following the approach detailed in Ref. \cite{jacksonMagnetoOpticalTrapField2019}.
One of the horizontal MOT beams was converted to linear polarization and scanned both horizontally and vertically across the MOT region while being kept on resonance with the $4s^2\, ^1S_0$ – $4s4p\, ^1P_1$ transition.
Its linear polarization was set parallel to the imaging camera, such that in the absence of any magnetic field, only a negligible amount of fluorescence light was emitted in the direction of the camera because of the usual $\sin^2(\theta)$ emission diagram of the $J=1, M_J=0 \rightarrow J=0, M_J=0$ fluorescing transition.
When the magnetic field gradient of the MOT is switched on, the nonzero magnetic field washes out this effect, and fluorescent light is detected by the camera, with the exception of the region where the magnetic field is zero, which then appears as a dark spot in the image, as shown in Fig. \ref{fig:MOT - zero field position}.
This allows to easily locate the zero-magnetic-field region in the vacuum chamber and to align the MOT beams accordingly.

The experimental parameters used to operate the MOT are summarized in Table \ref{table:MOT parameters}.

\begin{table*}[t!]
\centering
\renewcommand{\arraystretch}{2} % increases vertical spacing between rows
\setlength{\tabcolsep}{10pt}       % adjusts horizontal padding
\begin{tabular}{c c | c c}
\hline
\hline
\textbf{Oven temperature} & $723$ {K} & \textbf{Pressure in the MOT chamber} & $10^{-9}$ {mbar} \\
\hline
\textbf{MOT detuning} & $-36$ {MHz} & \textbf{Zeeman detuning} & $-600$ {MHz} \\
\hline
\makecell[c]{\textbf{MOT power}\\\textbf{in each arm}} & $2.13$, $2.30$, $0.687$ {mW} & \textbf{Zeeman power} & $32$ {mW} \\
\hline
\textbf{MOT waist} & $2$ {mm} & \textbf{Zeeman waist} & $1.5$ {mm} \\
\hline
\makecell[c]{\textbf{Saturation parameter}\\\textbf{in each MOT arm}} & $2.09$, $2.25$, $0.67$ & \makecell[c]{\textbf{Saturation parameter}\\\textbf{Zeeman}} & $15$ \\
\hline
$\boldsymbol{\nabla \vec{B}_{\mathrm{MOT}}}$ \textbf{axial} & 30 {G cm$^{-1}$} & $\boldsymbol{\nabla \vec{B}_{\mathrm{MOT}}}$ \textbf{radial} & $12$ {G cm$^{-1}$} \\
\hline
\hline
\end{tabular}
\caption{MOT and Zeeman slower operation parameters.}
\label{table:MOT parameters}
\end{table*}

\subsection{Electrode stack}
\label{subsec:electrode stack}
\begin{figure}[h!]
    \centering
    \includegraphics[width=0.4\textwidth]{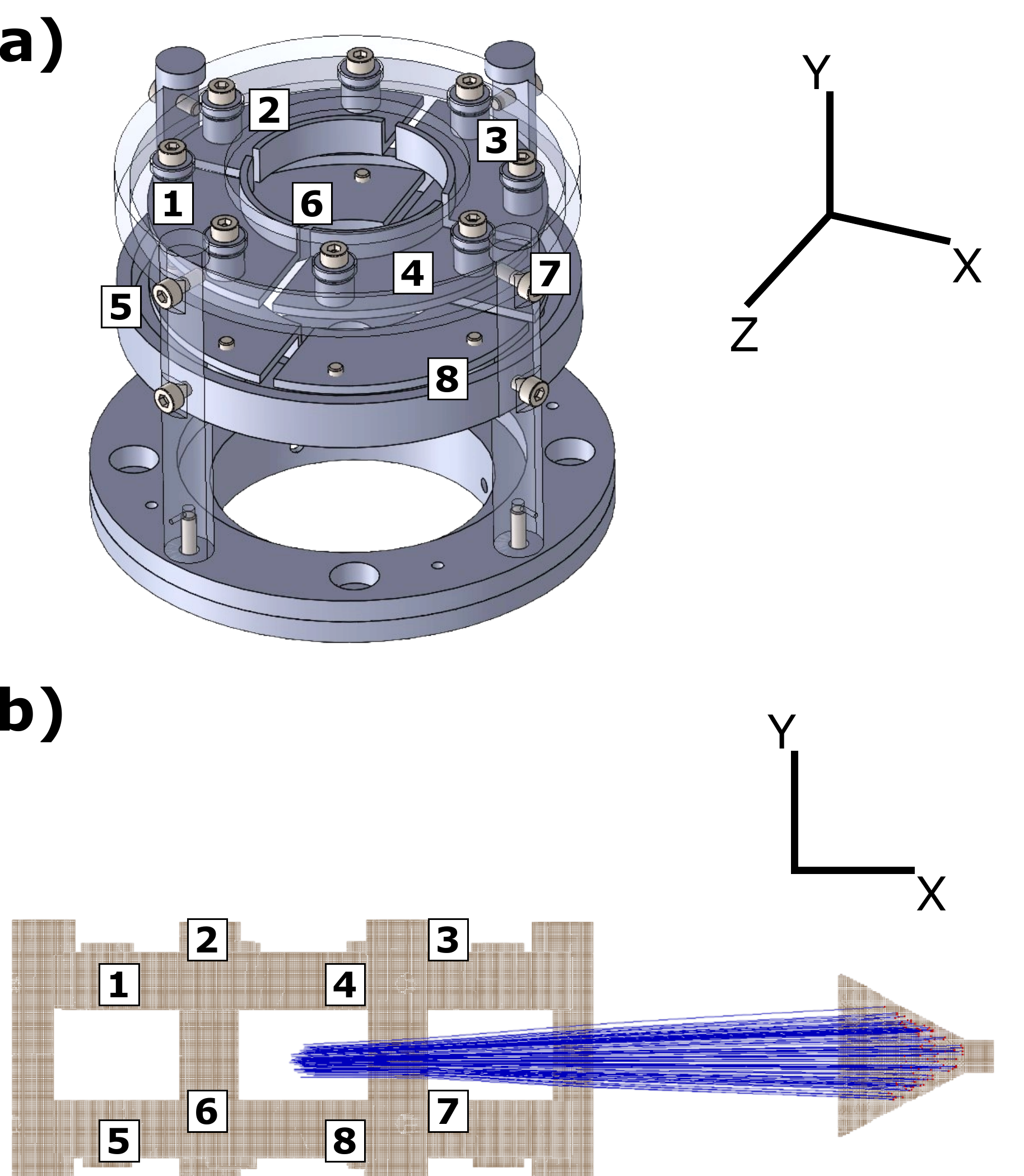}
    \caption{\textbf{a)} $3$D model of the electrode stack. Each electrode is numbered from $1$ to $8$. The electrodes $1$, $2$, $5$, and $6$ generate the extraction field that directs the ions toward the CEM. \textbf{b)} Simulation on SIMION \cite{dahlSimionPersonalComputer2000} of the trajectories of the ions generated inside the MOT when applying an extraction field to steer them into the CEM.}
    \label{fig:Electrodes}
\end{figure}

To apply electric fields of arbitrary direction in the MOT region, we designed and constructed an electrode-stack assembly consisting of two split rings made of stainless steel $316$ with internal and external diameters of $20$ and $40$ {mm}, respectively.
A schematic of the electrodes is presented in Fig. \ref{fig:Electrodes}\textbf{a}.
The rings, vertically stacked with $16$ {mm} separation, are each segmented into four independently addressable electrodes, allowing precise control of the electric-field orientation according to experimental requirements.

The electrodes generate electric fields that enable the pulsed-field ionization \cite{gallagherRydbergAtoms1994} of Rydberg atoms and the guiding of the resulting ions toward a channel electron multiplier (CEM, also called channeltron) for their detection.
Pulsed voltages of up to $3$ {kV} can be applied to the electrodes, which allows the field ionization of Rydberg states down to $n \sim 26$.
Figure \ref{fig:Electrodes}\textbf{b} shows a simulation of ion trajectories when voltages of $+2$ {kV} are applied on the four electrodes opposite to the CEM, corresponding to an electric field magnitude of $559$ {V cm$^{-1}$} in the MOT region.
All ions are efficiently steered onto the CEM.

A further application of the electrodes is the compensation of stray electric fields within the MOT region, since the polarizability of the Rydberg atoms, scaling as $n^7\,$ \cite{gallagherRydbergAtoms1994}, makes them extremely sensitive to even weak residual fields.
The segmentation of the electrodes means that stray fields can be compensated in all three directions of space. 

Finally, the electric fields generated by the electrodes may also be used for coupling the different states within the Stark manifold, enabling the excitation of high-$\ell$ Rydberg states via the Stark switching technique \cite{signolesCoherentTransferLowAngularMomentum2017}.

\section{Results}
\label{sec:results}
\begin{figure}[h!]
    \centering
    \includegraphics[width=0.5\textwidth]{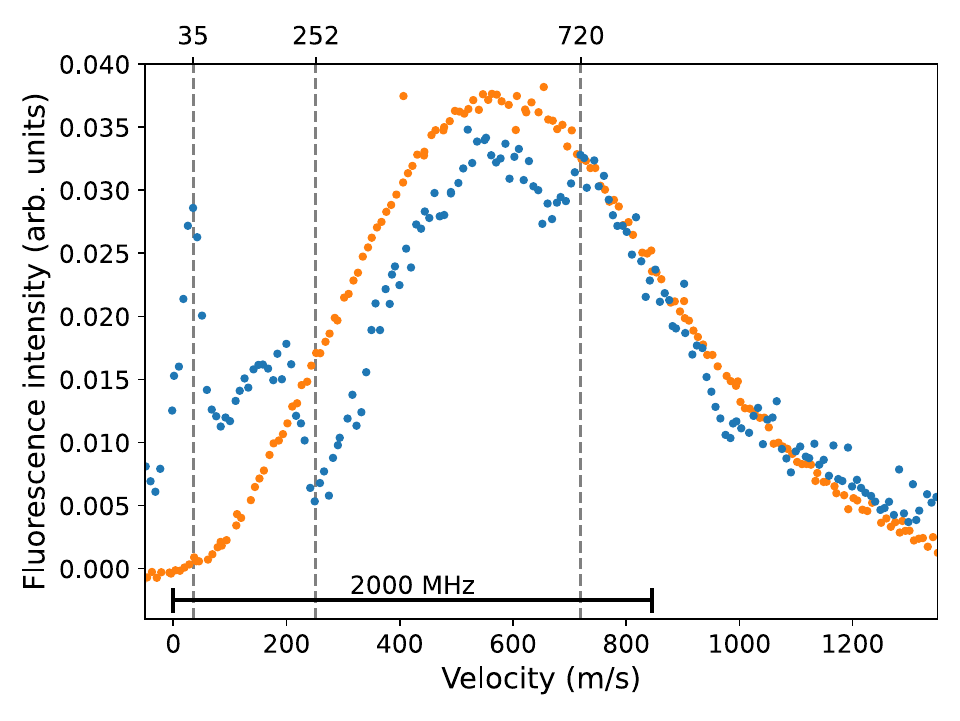}
    \caption{Velocity distribution of the atoms measured after the Zeeman slower without the slowing beam (orange full circles), and with the slowing beam (blue full circles).}
    \label{fig:Zeeman - polarization}
\end{figure}

The velocity distribution of the atoms was measured after the Zeeman slower, both with and without the slowing laser, by shining a second $423$-{nm} probe laser in the MOT region at a $\sim 45^\circ$ angle with respect to the atomic beam.
The intensity of the atomic fluorescence, recorded with a photomultiplier tube (PMT) while scanning the probe-laser frequency, is presented in Fig. \ref{fig:Zeeman - polarization}.

When the slowing laser beam is switched on, the atomic velocity distribution is modified in two ways. First, a depletion at around $250$ {m/s} is observed, accompanied by a redistribution of the population toward adjacent lower velocities.
This indicates that atoms initially in resonance with the slowing beam, red-detuned by $600$ {MHz}, are slowed down until they eventually fall out of resonance, which is consistent with a slowing process without magnetic compensation.
Second, a sharper peak around $35$ {m/s}, which demonstrates Zeeman slowing.
The maximum entrance velocity at which the atoms are slowed is $720$ {m/s}, in accordance with the simulated trajectories shown in Fig. \ref{fig:zeeman trajectories}. 
The final velocity of the atoms is about $35$ {m/s}, in good agreement with the design and calculated exit velocities, and well below the capture velocity of the MOT of $60$ {m/s}.

The slowing-beam diameter ($2w_0 = 3$ {mm}) is significantly smaller than the diameter of the Zeeman-slower vacuum pipe ($15$ {mm}).
Because the atomic beam in the oven chamber is $\sim 20$ {mm} and taking into account its transversal velocity spread, the atomic-beam size in the MOT chamber is geometrically limited by the Zeeman-slower vacuum pipe.
The relative fraction of slowed atoms, obtained by integrating the area of the Zeeman peak and comparing it to the total area of the velocity distribution below the cut-off velocity, is approximately $20\%$, in agreement with the ratio between the slowing-laser and atomic-beam radii, which is also $20\%$.

\begin{figure}[h!]
    \centering
    \includegraphics[width=0.5\textwidth]{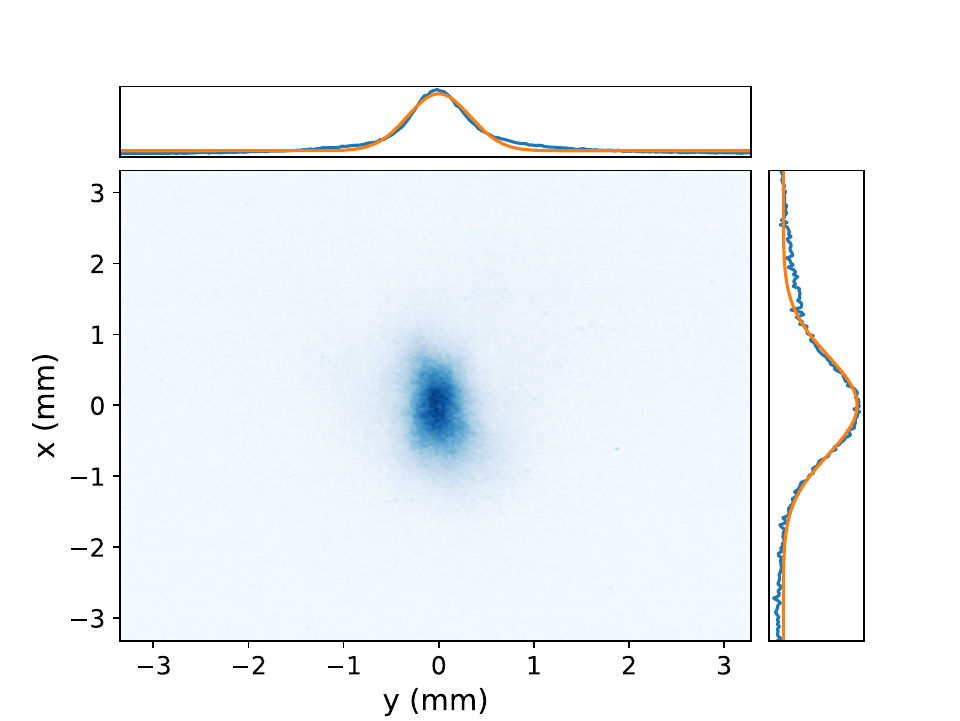}
    \caption{False-color fluorescence image of the MOT captured with the side-view camera, showing the signal projections integrated along the X and Y axes (blue solid lines) and the Gaussian fitting curves (orange solid lines).}
    \label{fig:MOT - fluorescence}
\end{figure}

The slow atoms exiting the Zeeman slower are then captured in the MOT. The atomic cloud size, the MOT trapping time, and the number and density of trapped atoms were determined by monitoring with the $423$-{nm} fluorescence of the cold atoms with both CMOS cameras and a PMT.
A typical fluorescence image of the MOT is shown in Fig. \ref{fig:MOT - fluorescence}.
The fluorescence intensity distribution closely resembles a Gaussian distribution, and the MOT size defined as the full width at half maximum (FWHM) of the Gaussian fit to the measured intensity profile, is  $0.83(1)$ {mm} and  $1.90(3)$ {mm} along the X and Y axes, respectively.
The elongated shape in the vertical direction, rather than a spherical distribution, arises from the lower optical power in the corresponding vertical MOT beams (see Table \ref{table:MOT parameters}).

To determine the number of trapped atoms, the optical power of the fluorescence light emitted through the solid angle defined by the surface of the imaging viewport was measured using a calibrated, high-sensitivity photodiode.
From the measured power of $4$ {nW}, we estimate the atom number to be approximately $10^6$, corresponding to a density of $10^8$ {cm$^{-3}$}.
This value represents a lower bound to the real atom number since the atoms in the metastable $4s3d\, ^1D_2$ and $4s4p\, ^3P_1$ do not fluoresce rapidly nor at the $423$-{nm} wavelength of our detection system, and therefore did not contribute to the measured signal.

Previous studies have shown that adding lasers at $672$ or $405$ nm enhances the MOT fluorescence signal by a factor of approximately $13\,$ \cite{burrowsRepumpMethodsMagneto2012} and $20\,$ \cite{millsEfficientRepumpingCa2017}, respectively, and consequently the number of trapped atoms.
These wavelengths repump the population from the $4s3d\, ^1D_2$ state to the $4s5p\, ^1P_1$ and the $4s8p\, ^1P_1$ states, respectively, thereby preventing decay into the long-lived $4s4p\, ^3P_2$ metastable state, which, with a lifetime of $118$ {min} \cite{dereviankoFeasibilityCoolingTrapping2001}, constitutes the main loss channel of the MOT.
Such repumpers not only increase the number of atoms but also extend the MOT trapping time, by a factor of three to nearly six in the case of the $672$-nm laser, as reported in \cite{burrowsRepumpMethodsMagneto2012, millsEfficientRepumpingCa2017}.
Other repump schemes have also been investigated in \cite{millsEfficientRepumpingCa2017}, identifying $453$ {nm} as the optimal wavelength for maximizing both the number of trapped atoms and the trapping time, with values up to $2.5$ {s} reported.
Their implementation in future iterations of the experiment is considered.

The temperature of the Ca atoms in the MOT was determined from the ballistic expansion of the atomic cloud. 
After loading the MOT for $140$ {ms}, the laser beams were switched off for a variable duration between $0$ and $5$ {ms}, after which they were switched back on. 
The reactivation of the MOT beams was synchronized with the trigger of the side-view camera to image the MOT cloud at different expansion times. 
Because the atomic cloud undergoes partial recompression when the MOT beams are turned back on, images were acquired with exposure times of $0.25$, $0.5$, $0.75$, and $1$ {ms} for comparison.

\begin{figure}[h!]
    \centering
    \includegraphics[width=0.5\textwidth]{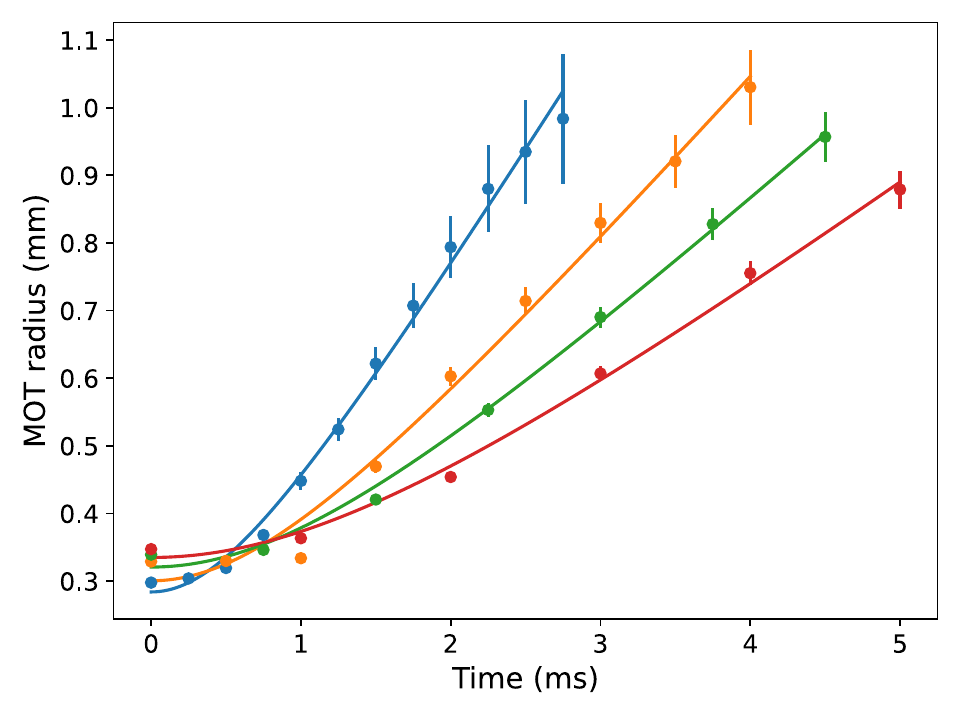}
    \caption{Ballistic expansion of the MOT cloud (solid circles) and fits to eq. \ref{eq:ballistic expansion} (solid lines). The data recorded at $0.25$, $0.5$, $0.75$, and $1$ {ms} exposure time is represented in blue, orange, green, and red, respectively.}
    \label{fig:ballistic expansion}
\end{figure}

The experimental results are presented in Fig. \ref{fig:ballistic expansion}.
Experimental data were fit to an equation describing the increase of the RMS radius of the atomic cloud $r$ with the duration of the ballistic expansion $t$ (see e.g. Ref. \cite{sukachevSubdopplerLaserCooling2010}).
To account for the recompression, an additional factor was added to the $k_B Tt^2/m$ term to account for the overdamped motion of the atoms when the MOT beams are switched back on and during the exposure time $t_{exp}$ necessary to acquire the image.
The formula is then

\begin{equation}
r(t) = \sqrt{r_0^2 + \frac{k_B T}{m} t^2 e^{-(\frac{t_{exp}}{\tau_d})^2}}
\label{eq:ballistic expansion}
\end{equation}

where $r_0$ denotes the radius of the atomic cloud, defined as the standard deviation of the corresponding normal distribution before ballistic expansion; $k_B$ is the Boltzmann constant; $T$ is the MOT temperature; $m$ is the mass of $^{40}$Ca; $t$ is the delay time between switching off the MOT and triggering the camera; $t_{exp}$ is the camera exposure time; and $\tau_d$ is the characteristic damping time of the MOT.
The latter was calculating using $\tau_d = 2\Gamma_\text{MOT}/\omega_\text{MOT}^2$, with $\Gamma_\text{MOT}$ the MOT damping rate ($\sim 43$ {kHz} in the present case) and $\omega_\text{MOT}$ the MOT trapping frequency ($\sim 7$ {kHz}).
Both can be calculated from standard formulas \cite{metcalfLaserCoolingTrapping1999} using the experimental parameters we measured and listed in table \ref{table:MOT parameters}.

The temperature obtained from fitting the different data sets is $1.0(3)$ {mK}, in agreement with the value reported in \cite{adamczykTwophotonCoolingCalcium2025} and consistent with the Doppler limit of $0.8$ {mK}.
Small variations in the initial MOT radius, $r_0$, of about $50$ {$\mu$m} can be observed.
The fits were repeated after averaging and fixing the value of $r_0$, leaving only the temperature as a free parameter, and yielded the same result, demonstrating that these small fluctuations represent drifts of the MOT cloud size over long periods of time, and not fluctuations of the temperature.

Finally, the loading time of the MOT was measured by rapidly switching on the MOT laser beams with an acousto-optic modulator and monitoring the increase of fluorescence in time with the PMT.
With a constant flux of atoms arriving in the MOT chamber from the Zeeman slower, turning on the MOT laser beams causes the slow atoms entering the trapping region to be cooled and confined.
As a result, the MOT fluorescence increases as the atomic density grows.
The evolution of the population trapped in the MOT over time is given by Eq. \ref{eq:population dynamics} \cite{arpornthipVacuumpressureMeasurementUsing2012},

\begin{equation}
\frac{\mathrm{d}N}{\mathrm{d}t} = R - \gamma N(t) - \beta \bar{n} N
\label{eq:population dynamics}
\end{equation}

where $R$ is the loading rate of the MOT, $\gamma = \gamma_{\textrm{col}} + \gamma_{\textrm{met}}$ is the rate of one-atom loss events, which includes collisions with the background gas ($\gamma_{\textrm{col}}$) and decay to metastable states ($\gamma_{\textrm{met}}$). $N$ denotes the atom number, $\beta$ is the two-body loss rate corresponding to inelastic collisions between trapped atoms, and $\bar{n}$ is the average atomic density within the MOT.

At relatively low optical power and atomic density, typically $\gamma > \beta \bar{n}$ and the evolution of the number of atoms N when loading the MOT is

\begin{equation}
N(t) = N_\infty (1 - \exp{-t / \tau})
\end{equation}

with $N_\infty \simeq 10^{6}$ and $\tau$ being the loading time of the MOT.
In the steady state, the loading rate is equal to the loss rate, such that the loading time and trapping (decay) time are identical.
% the evolution of the population trapped in the MOT can be described by an exponential curve. In this regime, the fluorescence signal increases exponentially until the rate of atoms captured equals the total loss rate, reaching a steady-state population. Under these conditions, the loading time and trapping (decay) time are identical.

\begin{figure}[h!]
    \centering
    \includegraphics[width=0.5\textwidth]{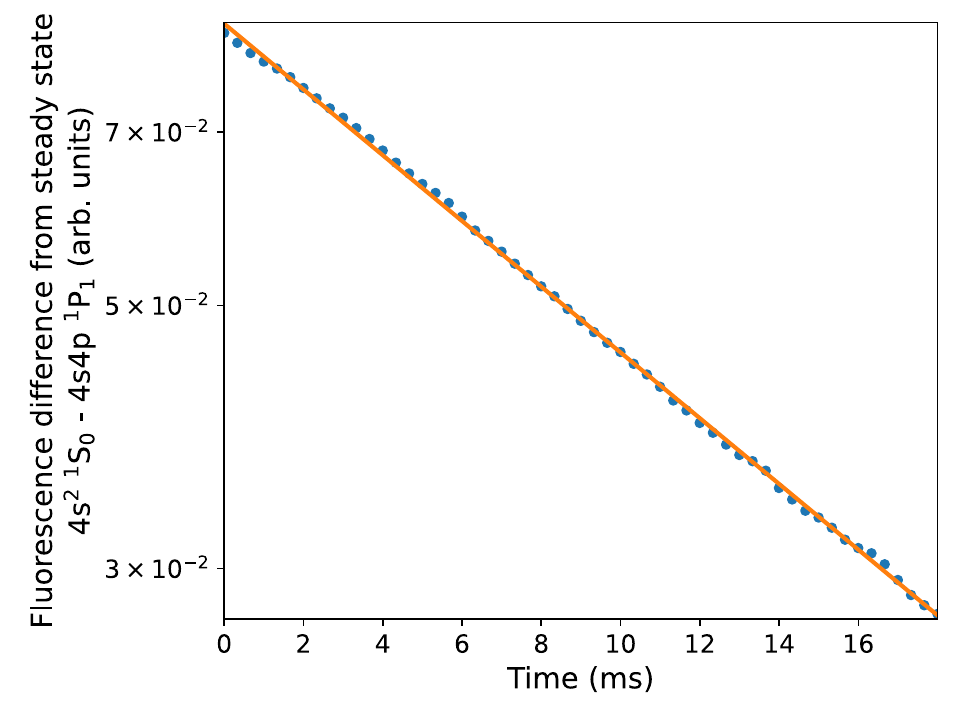}
    \caption{Difference between the MOT fluorescence at a given time during loading and its steady-state level on a logarithmic scale. The solid circles represent the experimental measurements, and the solid line, the exponential fitting yielding a trapping time of $16(2)$ {ms}.}
    \label{fig:MOT - trapping time}
\end{figure}

Figure \ref{fig:MOT - trapping time} shows $N_\infty - N(t)$, together with a fit to the function $e^{-t/\tau}$ that yields a loading time of $16(2)$ {ms}.
This value is consistent with previous reports for calcium MOTs operated under similar conditions without a repumping system \cite{burrowsRepumpMethodsMagneto2012, cavassofilhoCalciumMagnetoopticalTrap2003}, and it corresponds to the limit set by the decay into the metastable state $4s4p\, ^3P_2$.

\begin{figure}[h!]
    \centering
    \includegraphics[width=0.5\textwidth]{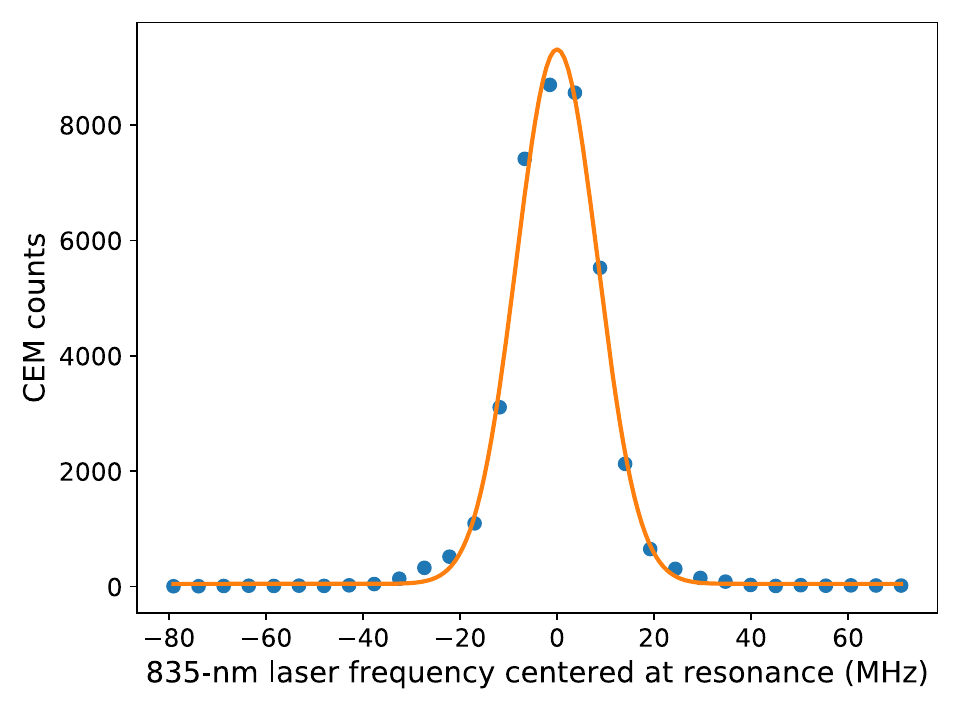}
    \caption{Rydberg spectrum obtained by pulsed-field ionization of the $4s54p,{}^1P_1$ Rydberg state while scanning the $835$-{nm} laser frequency. The blue solid circles and the orange line correspond to the experimental values and the Gaussian fitting, respectively.}
    \label{fig:Rydberg spectrum}
\end{figure}

The ground-state Ca atoms in the MOT can then be excited to high Rydberg states by means of resonant three-photon excitation.
The first transition ($4s^2\, ^1S_0$ – $4s4p\, ^1P_1$) is driven by the MOT lasers.
The second ($4s4p\, ^1P_1$ - $4s4d\, ^1D_2$, $732$ {nm}) and third transitions ($4s4d\, ^1D_2$ - $4snp\, ^1P_1$ or $4snf\, ^1F_3$, $835$ {nm}) are driven by commercial external-cavity diode lasers whose frequencies are stabilized to an accurate wavemeter.
Figure \ref{fig:Rydberg spectrum} shows a typical excitation spectrum to the $4s54p\, ^1P_1$ Rydberg level, recorded after letting the second and third lasers interact with the atoms in the MOT for $\sim 100$ $\mu$s.
An electric field of $2$ {kV} is then applied with a fast high-voltage switch to the four electrodes on the side opposite to the CEM.
The electric field ionizes the Rydberg atoms and accelerates the Ca$^+$ onto the CEM, where they are detected and counted with fast electronics.
With this setup, we have recorded $4snp$ and $4snf$ Rydberg states from $n \sim 26$ up to $n > 150$, at excitation frequencies in excellent agreement with Ref. \cite{miyabeDeterminationIonizationPotential2006}.

\section{Conclusions}
\label{sec:conclusions}
We have presented an experimental apparatus for laser cooling and trapping Ca atoms in an ultra-high-vacuum environment using the Ca($4s^2, ^1S_0$ – $4s4p, ^1P_1$) transition at $423$ {nm}.
Ground-state atoms are then excited to high Rydberg states with a three-photon resonant-excitation scheme, and an electrode-stack system surrounding the trapping volume enables the application of electric fields, the field ionization of the Rydberg atoms, and the detection of the resulting positive ions.
In the first stage of the experiment, the Ca atoms are evaporated in an oven, and emitted in a beam with $0.11$ {rad} divergence at a most probable velocity of $650$ {m/s}.
The atoms are then slowed down to a velocity of $35$ {m/s} by means of a permanent-magnet-based Zeeman slower.
Subsequently, the slow atoms are captured in a magneto-optical trap, where they remain confined for $16$ {ms} on average, while they lose energy until reaching temperatures around $1$ {mK} with a density of approximately $10^8$ {cm$^{-3}$}.
These results are consistent with previous reports of magneto-optical traps of similar characteristics \cite{adamczykTwophotonCoolingCalcium2025, cavassofilhoCalciumMagnetoopticalTrap2003, burrowsRepumpMethodsMagneto2012}.
Once the atoms are cooled and trapped in the MOT, the electrode stack can be used to apply electric fields to, e.g., control Rydberg excitation via Stark switching, as well as for pulsed-field ionization and detection of the Rydberg states.
The results discussed above confirm that the setup described in the present paper provides a robust platform for experiments with ultracold Ca Rydberg atoms.

\section{Acknowledgments}
\label{sec:acknowledgements}
This work was supported by the Fonds de la Recherche Scientifique—FNRS under MIS Grant No. F.4027.24 and IISN Grant No. 4.4504.10, and by the Secuweb project (0100085) of the Interreg France-Wallonia-Flanders programme, an initiative co-financed by the European Union (European Regional Development Fund).

E.M.B. is financially supported by a FRIA fellowship from the Fonds de la Recherche Scientifique – FNRS.

X.U. is a Senior Research Associate of the Fonds de la Recherche Scientifique – FNRS

\bibliography{bibliography}

% If in two-column mode, this environment will change to single-column format so that long equations can be displayed. 
% Use only when necessary.
%\begin{widetext}
%$$\mbox{put long equation here}$$
%\end{widetext}

% Figures should be put into the text as floats. 
% Use the graphics or graphicx packages (distributed with LaTeX2e).
% See the LaTeX Graphics Companion by Michel Goosens, Sebastian Rahtz, and Frank Mittelbach for examples. 
%
% Here is an example of the general form of a figure:
% Fill in the caption in the braces of the \caption{} command. 
% Put the label that you will use with \ref{} command in the braces of the \label{} command.
%
% \begin{figure}
% \includegraphics{}%
% \caption{\label{}}%
% \end{figure}

% Tables may be be put in the text as floats.
% Here is an example of the general form of a table:
% Fill in the caption in the braces of the \caption{} command. Put the label
% that you will use with \ref{} command in the braces of the \label{} command.
% Insert the column specifiers (l, r, c, d, etc.) in the empty braces of the
% \begin{tabular}{} command.
%
% \begin{table}
% \caption{\label{} }
% \begin{tabular}{}
% \end{tabular}
% \end{table}

% If you have acknowledgments, this puts in the proper section head.
%\begin{acknowledgments}
% Put your acknowledgments here.
%\end{acknowledgments}

% Create the reference section using BibTeX:
% \bibliography{your-bib-file}

\end{document}